\documentclass[twocolumn,aps,amsmath,amssymb,showpacs]{revtex4}

\def\al{\alpha}
\def\be{\beta}

\def\la{\lambda}

\def\la{\lambda}
\def\vp{\varphi}

\def\E{{\rm E}_{10}}
\def\KE{{{\rm K}(\E})}

\def\bfa{{\boldsymbol{{\tt a}}}}
\def\bfb{{\boldsymbol{{\tt b}}}}
\def\bbfa{\bar{{\boldsymbol{{\tt a}}}}}

\def\btau{{\boldsymbol\tau}}
\def\bom{{\boldsymbol\omega}}

\def\cI{{\cal I}}

\def\bi{{\bf a}}
\def\bj{{\bf b}}
\def\bk{{\bf c}}

\def\bE{{\bf 1}}
\def\bZ{{\bf 2}}
\def\bD{{\bf 3}}
\def\bV{{\bf 4}}
\def\bS{{\bf 6}}
\def\bA{{\bf 8}}

\def\sa{\mathfrak{a}}
\def\sb{\mathfrak{b}}

\def\bsa{\bar{\mathfrak{a}}}
\def\bsb{\bar{\mathfrak{b}}}

\def\bbE{{\bar{\bf 1}}}
\def\bbZ{{\bar{\bf 2}}}
\def\bbD{{\bar{\bf 3}}}
\def\bbV{{\bar{\bf 4}}}
\def\bbS{{\bar{\bf 6}}}

\def\rS{{\rm S}}
\def\rG{{\rm G}}

\def\bbi{{\bar{\bf a}}}
\def\bbj{{\bar{\bf b}}}
\def\bbk{{\bar{\bf c}}}

\def\mG{\mathfrak{G}}
\def\mN{\mathfrak{N}}

\def\eq{\!=\!}

\def\cR{{\mathcal R}}
\def\cU{{\mathcal U}}
\def\11{{\mathbb 1}}

\def\ZZ{{\mathbb Z}}

\def\ri{{\rm i}}

\def\beq{\begin{equation}}
\def\eeq{\end{equation}}
\def\bea{\begin{eqnarray}}
\def\eea{\end{eqnarray}}
\def\nn{\nonumber}

\def\ra{\rightarrow}

\def\ri{\text{i}}

\begin{document}
\title{Standard Model Fermions and Infinite-Dimensional R-Symmetries}
\author{Krzysztof A. Meissner$^{1,2}$ and Hermann Nicolai$^1$}
\affiliation{$^1$ Max-Planck-Institut f\"ur Gravitationsphysik
(Albert-Einstein-Institut)\\
M\"uhlenberg 1, D-14476 Potsdam, Germany\\
$^2$ Institute of Theoretical Physics, Faculty of Physics,
University of Warsaw\\
Pasteura 5, 02-093 Warsaw, Poland}

\vspace{3mm}

\begin{abstract} Following up on our earlier work \cite{MN} where we showed how to
amend a scheme originally proposed by M. Gell-Mann to identify the 48 spin-$\frac12$
fermions of $N\eq 8$ supergravity that remain after complete breaking of $N\eq 8$ 
supersymmetry with the 3$\,\times\,$16 quarks and leptons of the Standard Model, 
we further generalize the construction to account for the full 
SU(3)$_c\,\times\,$SU(2)$_w\,\times\,$U(1)$_Y$
assignments, with an additional family symmetry SU(3)$_f$.
Our proposal relies in an essential way on embedding the SU(8) 
R-symmetry of $N\eq 8$ supergravity into the (infinite-dimensional)
`maximal compact' subgroup $\KE$ of the conjectured maximal duality symmetry $\E$. 
As a by-product, it predicts fractionally charged and possibly strongly interacting 
massive gravitinos. It also indicates how 
$\E$  and $\KE$ can supersede supersymmetry as a guiding principle for unification.
\end{abstract}
\maketitle

It is generally believed that neither the particle content nor the dynamics of maximal 
$N\!=\!8$ supergravity \cite{CJ,dWN} can be matched with the Standard Model (SM) 
of particle physics. In this Letter we re-examine this widely accepted wisdom on the basis 
of recent work \cite{MN,KN} which further develops an old proposal of Gell-Mann's 
\cite{GellMann} to identify the 48 quarks and leptons of the SM with the 48 spin-$\frac12$  
fermions that remain after complete breaking of $N\eq 8$ supersymmetry.
As shown in \cite{MN} the mismatch $\pm \frac16$  of electric charges in the original 
Gell-Mann scheme can be rectified by introducing a new vectorlike generator $\cI$
not contained in the SU(8) R-symmetry group of $N\eq 8$ supergravity.
However, in subsequent work \cite{KN} it was shown that this new generator does belong
to $\KE$, the maximal compact subgroup of the hyperbolic Kac--Moody group $\E$ 
conjectured to be a symmetry underlying M theory \cite{DHN,West}, 
thus demonstrating the need for new ingredients {\em beyond} $N\!=\!8$ supergravity.    
Here we take this construction one step further by exploring how 
the full SU(3)$_c\, \times$ SU(2)$_w \, \times$ U(1)$_Y$ symmetry of the SM might be 
incorporated into this scheme, together with an additional family symmetry SU(3)$_f$ that
does not commute with the electroweak symmetries.

As the present proposal differs substantially from current paradigms to 
derive the SM fermions from a Planck scale unified theory of quantum gravity,  
let us first explain our basic `philosophy'. Our 
considerations here are based on the central conjecture of \cite{DHN}
according to which the full $\E$ symmetry conjectured to underlie M theory manifests itself 
in a `near singularity limit'. This conjecture itself is based on a BKL-type analysis of 
cosmological singularities, where the causal decoupling of spatial points entails 
an effective dimensional reduction to one (time) dimension \cite{BKL}.
In this pre-geometric regime, the spatial dependence of the fields is supposed 
to `de-emerge' near the singularity, in the sense that it gets `spread' over the Lie 
algebra of $\E$, as outlined in \cite{DHN}. We emphasize, however, that 
explaining the emergence of a space-time based quantum field theory in this scheme,
as well as explaining the emergence of space-time symmetries and the distinction between 
global and local symmetries, is still an unsolved problem.

Subsequent analysis of the fermionic sector of the theory \cite{DKN} has revealed that the relevant 
symmetry acting on the fermions is the involutory (`maximal compact')  subgroup $\KE$ of 
$\E$, the conjectured {\em R-symmetry of M theory}. Here as well, one restricts 
attention in a first step to the fermionic fields {\em at a given spatial point}, with the idea 
that the spatial dependence would emerge in a more complete description
based on more faithful realizations of $\KE$. So far, however,  only finite-dimensional 
{\em unfaithful} spinorial (double-valued) representations of $\KE$ are known \cite{DKN,dBHP},
the two simplest of which are in one-to-one correspondence with the spinors 
of maximal supergravity. The action of $\KE$ on these representations can 
be conveniently described in 
terms of the quotient group $\mG_\cR = \KE/\mN_\cR$ where $\mN_\cR$ is the normal subgroup 
associated with the ideal spanned by those $\KE$ generators that 
annihilate all elements of the given representation space $\cR$. For the Dirac 
representation corresponding to the supersymmetry parameter we 
have $\mG_D =$ Spin(32)$/\ZZ_2$, while for the Rarita-Schwinger  
representation describing the physical fermionic degrees of freedom we have 
$\mG_{RS}=$ Spin(288,32)$/\ZZ_2$ \cite{Vigano}; the latter group also contains
transformations acting chirally on the $N\eq 8$ fermions. We expect there to 
exist bigger and less unfaithful representations which can possibly capture more 
of the spatial dependence \cite{KN1}.

Although the present scheme necessarily transcends $N\eq 8$ supergravity, this 
theory nevertheless continues to play a crucial role in `guiding' our proposal, in that
we insist on remaining compatible with the original scheme of \cite{GellMann}, and with
the known vacuum structure of gauged maximal supergravities with stationary 
points preserving a residual SU(3)$\,\times\,$U(1) symmetry \cite{Warner,NW,Hull,Dall}.
The group SU(3)$_c\,\times\,$U(1)$_{em}$ 
is believed to be the gauge symmetry that survives to the lowest energies in the SM, 
but a naive identification of the supergravity SU(3) with the color group SU(3)$_c$ 
does not work, as is immediately obvious from (\ref{Fermions1}) below.
For this reason, M.~Gell-Mann introduced an additional family symmetry SU(3)$_f$ that 
acts between the three particle families (generations) and proposed to identify
the residual SU(3) of supergravity with the {\em diagonal subgroup} of color and 
family \cite{GellMann}. This scheme `almost'  works in the sense that, after the 
removal of eight Goldstinos there is complete agreement of the SU(3) assignments, 
but there remains a systematic mismatch in that the U(1) charges of the 
supergravity fermions are systemically off by $\pm \frac16$ from the electric charges 
of the quarks and leptons. As noticed in \cite{MN} this mismatch can be remedied  
by means of the extra generator (\ref{cI}) which `deforms' the SU(3)$\,\times\,$U(1) group. 
Although this deformation is no longer contained in SU(8), it {\em is} contained  
in $\KE$ \cite{KN}. 
One notable feature here is that even though accompanied by 
a vast enlargement of the R-symmetry, our proposal makes do with the original 
56 spin-$\frac12$ fermions and eight gravitinos, whereas more conventional 
schemes would require correspondingly larger multiplets for larger groups.

Accordingly, we now focus on the fermionic sector of $N\eq 8$
supergravity, which consists of eight gravitinos $\psi_\mu^i$ transforming in the $\bA$, 
and a tri-spinor of spin-$\frac12$ fermions $\chi^{ijk}$ transforming in the $\bf{56}$ of SU(8), 
where  $\chi^{ijk}$ is fully antisymmetric in the SU(8) indices $i,j,k$. We adopt the conventions 
and notations of \cite{dWN}, where complex conjugation raises (or lowers) indices, such that
for instance $\chi^{ijk} = \big( \chi_{ijk} \big)^*$, and where upper (lower)  position
of the SU(8) indices indicates positive (negative) chirality.
Hence the chiral SU(8) transformations act as
$\chi^{ijk}  \ra  U^i{}_l U^j{}_m U^k{}_n \chi^{lmn} \, , \;
\chi_{ijk}  \ra U_i{}^l U_j{}^m U_k{}^n \chi_{lmn}
$
with $U \in$ SU(8), $U_i{}^j \equiv \big( U^i{}_j \big)^*$ and $U^i{}_k U_j{}^k = \delta^i_j$.

As already pointed out, a special role is played by the  subgroup SU(3)$\,\times\,$U(1), which 
is contained in the vectorlike gauge group SO(8) $\subset$ SU(8) via the real embedding
SU(3) $\times$ U(1) $\subset$ SO(6)$\,\times\,$SO(2) $\subset$ SO(8) \cite{GellMann,NW}. 
To study the relevant decompositions we introduce boldface indices $\bE,...,\bV$, 
and barred indices $\bbE,...,\bbV$ for the conjugate representations as in \cite{NW,MN},
but now with unit normalization so that $v^\bE\equiv 2^{-1/2} (v^1 + \ri v^2), v^\bZ \equiv
2^{-1/2}(v^3 + \ri v^4)$, {\it etc.} and $v^\bbE \equiv 2^{-1/2}(v^1 - \ri v^2), v^\bbZ \equiv
2^{-1/2}(v^3 - \ri v^4)$, {\it etc.}
Similar rules apply to the fermions; for instance,
\bea\label{vp}
\vp^{\bE\bZ\bbV} &\equiv& \frac1{2\sqrt{2}} \Big(\chi^{137} + \ri\chi^{237} + \ri\chi^{147} - \ri\chi^{138} \nn\\
          && \qquad -\chi^{247} + \chi^{238} + \chi^{148} + \ri \chi^{248}\Big) \, ,
\eea          
and so on. It is important here that the spinor $\vp^{\bbE\bbZ\bV}$ with barred and 
unbarred indices interchanged is {\em not}  the complex conjugate of $\vp^{\bE\bZ\bbV}$ 
because the spinor components $\chi^{ijk}$ are themselves complex. Writing (\ref{vp}) in 
the schematic form $\vp = \rS\circ\chi$ it is easy to see that $\rS$ decomposes into 
12 blocks of 2-by-2 matrices and four blocks of 8-by-8 matrices, see \cite{Link}
for explicit formulas.

For the further analysis we single out the first three indices with labels 
$\bi ,\bj, \dots= \bE,\bZ,\bD$, on which the SU(3) subgroup acts. 
There is furthermore a two-parameter family of U(1) subgroups
embedded as follows into SO(8)
\begin{equation} \label{U(1)}
 Y(\al,\be) \,= \, \begin{pmatrix}
                0 & - \al  & 0 & 0 & 0 & 0 & 0 & 0 \\
                \al  & 0 & 0 & 0 & 0 & 0 & 0 & 0 \\
                0 & 0 & 0 & -\al & 0 & 0 & 0 & 0 \\
                0 & 0 & \al & 0 & 0 & 0 & 0 & 0 \\
                0 & 0 & 0 & 0 & 0 & - \al & 0 & 0 \\
                0 & 0 & 0 & 0 & \al & 0 & 0 & 0 \\
                0 & 0 & 0 & 0 & 0 & 0 & 0 & -\beta\\
                0 & 0 & 0 & 0 & 0 & 0 & \beta & 0
                \end{pmatrix},
\end{equation}
This matrix commutes with U(3)$\,\times\,$U(1) $\subset$ SO(8) for all $\al, \be$, and thus defines 
an SU(3)$\,\times\,$U(1) subgroup of SO(8) for each choice of $\al$ and $\be$.
As shown in \cite{GellMann} we must take
\beq\label{albe}
\al = \frac16 \;\; , \quad\be = \frac12 \, ,
\eeq
in order to match the $N\!=\!8$ fermions with those of the Standard Model (and incidentally
also in accord with the structure of $N\!=\!2$ AdS supermultiplets \cite{NW}). 
With this choice one easily reads off the SU(3)$\,\times\,$U(1) assignments for the gravitinos
\bea\label{Gravitino1}
\psi_\mu^\bi &\in&  (\bD\,,\, {\small\frac16)}\; , \quad \psi_\mu^\bbi \in (\bbD\,,\, - \frac16)\; ,  \nn\\[2mm]
\psi_\mu^\bV &\in& (\bE \,,\,\frac12) \; , \quad \psi_\mu^\bbV \in (\bE \,,\, - \frac12)
\eea
The 56 spin-$\frac12$ fermions are split into 6+2 Goldstinos
{\small
\bea\label{Goldstino}
\vp^\bi &\equiv& \vp^{\bi\bV\bbV} \,\in\, (\bD\,,\, \frac16) \; , \;\;
\vp^\bbi \equiv \vp^{\bbi\bV\bbV} \,\in \, (\bbD\,,\, - \frac16) \;\; , \nn\\[2mm]
\vp^\bV &\equiv& \vp^{\bi\bj\bk} \in (\bE\,,\, \frac12) \;,\;\;
\vp^\bbV \equiv \vp^{\bbi\bbj\bbk} \,\in \,(\bE\,,\, -\frac12)
\eea}
and the remaining 48 spin-$\frac12$ fermions:
{\small
\bea\label{Fermions1}
\vp^{\bi\bj\bV} &\in& (\bbD\,,\, \frac56) \; , \quad \vp^{\bi\bj\bbV} \in\, (\bbD\,,\, - \frac16)\,,\quad 
\\[1mm]
\vp^{\bbi\bbj\bV} &\in& (\bD\,,\, \frac16) \; , \quad \vp^{\bbi\bbj\bbV}\in (\bD\,,\, - \frac56) \nn
\\[1mm]
\vp^{\bi\bj\bbk} &\in&  (\bD,\frac16) \oplus (\bbS,\frac16) \;,\;
\vp^{\bbi\bbj\bk} \in (\bbD, - \frac16) \oplus (\bS\,,\,-\frac16) \;\;
\nn\\[1mm]
\vp^{\bi\bbj\bV} &\in& ({\bf 8}\,,\,\frac12) \oplus (\bE\, ,\,\frac12) \;,\;
\vp^{\bi\bbj\bbV} \in ({\bf 8}\,,\, - \frac12) \oplus (\bE\,,\, -\frac12)  \nn
\eea
}
To be sure, in order to properly identify the Goldstinos, we must allow for a possible mixing 
between the representations with the same SU(3)$\,\times\,$U(1) content, as is also obvious 
from the mass eigenstates at the SU(3)$\,\times\,$U(1) stationary point \cite{NW}.
However, the precise form of the mixing depends on the dynamics which is expected 
to deviate from $N\eq 8$ supergravity and which at this point is unknown. 
It is an essential assumption here that even with such mixing we can bring the fermions 
to the form with the labeling exhibited above, possibly by means of a chiral redefinition
of the fermions. We then assume that all supersymmetries are broken at a high 
scale, such that all gravitinos acquire a very large mass.

Now, as first pointed out in \cite{GellMann}, with the above choice of Goldstinos the three 
generations of SM fermions can be matched with the remaining 48 spin-$\frac12$ fermions 
from (\ref{Fermions1}), provided one identifies the  supergravity SU(3) with the diagonal 
subgroup of color SU(3)$_c$ and a new family symmetry SU(3)$_f$, {\it viz.}
\beq
{\rm SU}(3) \equiv \big[ {\rm SU}(3)_c \times {\rm SU}(3)_f \big]_{\rm diag}
\eeq
and furthermore allows for a spurion charge $\pm \frac16$ to correct the U(1)
charges in (\ref{Fermions1}) so as to recover the known electric charges of quarks 
and leptons. More specifically, the assignments are as follows \cite{GellMann}
{\small
\begin{align}
\label{Fermions2}
&(u,c,t)_L &   \, &\, \bD_c \times \bbD_f    \,\ra \, \bA \oplus \bE  & \
\frac23 &= \frac12 + q \nn\\
&(\bar{u}, \bar{c}, \bar{t})_L &  \, &\, \bbD_c \times \bD_f  \,\ra \, \bA \oplus \bE & \ - \frac23 & = -\frac12 - q\nn\\
&(d,s,b)_L & \, & \, \bD_c \times \bD_f   \, \ra \, \bS \oplus \bbD  &
\  - \frac13 & =  - \frac16 -  q \nn\\
&(\bar{d}, \bar{s}, \bar{b})_L &  \,& \, \bbD_c \times \bbD_f  \, \ra \,\bbS \oplus \bD & \ \frac13 & = \frac16 + q\nn\\
&(\nu_e,\nu_\mu,\nu_\tau)_L &  \, & \, \bE_c \times \bbD_f    \,\ra \,
\bbD  & \  0 &=  - \frac16 +  q \nn\\
&(\bar{\nu}_e,\bar{\nu}_\mu, \bar{\nu}_\tau)_L & \, & \, \bE_c \times \bD_f  \, \ra \, \bD & \ 
0 &= \frac16 - q\nn\\
&(e^-,\mu^-,\tau^-)_L &  \, & \, \bE_c \times \bD_f    \,\ra \,
\bD  & \ - 1 & =  - \frac56  -  q \nn\\
&(e^+,\mu^+,\tau^+)_L &  \,  & \, \bE_c \times \bbD_f  \, \ra \, \bbD
& \ 1  & =  \frac56  +  q
\end{align}
}
where we made use of the fact that right-chiral particles can be equivalently described 
by their left-chiral anti-particles, and where the spurion shift is $q=-\frac16$ (resp.
$q= +\frac16$) for triplets (resp. antitriplets) of SU(3)$_f$.

The main advance reported in \cite{MN} was to show that with the representation
(\ref{Fermions1}) of the $N\eq 8$ spin-$\frac12$ fermions the spurion shift can be 
accounted for by means of the U(1)$_q$ transformations $\exp\left(\frac16 \omega \cI\right)$, where
\beq\label{cI}
\cI := \frac12 \Big(T \otimes {\bf 1} \otimes {\bf 1} \, + \,  {\bf 1} \otimes T \otimes {\bf 1} \, + \,
        {\bf 1} \otimes {\bf 1} \otimes T \, + \,  T \otimes T \otimes T \Big)
\eeq
with $T \equiv Y(1,1)$ (cf. (\ref{U(1)})). 
Because $T$ acts as $+ \ri$ (or $-\ri$) on an unbarred (barred) index it is easy to see
that $\cI$ gives $(+\ri)$ on $\vp$'s with no or only one barred 
index, and $(-\ri)$ on $\vp$'s with two or three barred indices. Importantly, the term with 
$T\otimes T \otimes T$, and hence the generator $\cI$ are {\em not} elements of SU(8).
However, in \cite{KN} it was shown that this new generator does belong to $\KE$, and
that furthermore the action of the generator $T\otimes T \otimes T$ can be extended 
to the gravitinos, with $ \cI \circ \psi = T \circ \psi$. Therefore
the action of $\frac16 \cI$ adds $\pm\frac16$ to the U(1) charges of the gravitinos 
in (\ref{Gravitino1}). Consequently, $\psi_\mu^\bi$ has electric charge $\frac13$, and 
$\psi_\mu^\bV$ has electric charge $\frac23$, consistent with the assignments (\ref{Goldstino}); 
$\psi_\mu^\bbi$ and $\psi_\mu^\bbV$ have the opposite electric charges, 

While the new generator $\cI$ commutes with the original SU(3)$\,\times\,$U(1),
and therefore merely deforms it but does not enlarge it, this is not so with 
the remaining generators of SU(8). As shown in  \cite{Vigano}, repeated commutation 
of $\cI$ with the elements of SU(8) generates a much bigger group acting 
on $\chi^{ijk}$, namely SU(56). Because $\cI\in\KE$, the latter group is also contained 
in $\KE$, and should consequently be viewed as a subgroup of the quotient 
group $\mG_{RS} = $ Spin(288,32)$/\ZZ_2$. The fact that this SU(56) group acts chirally 
motivates us to look for a realization of the full SM symmetries within $\KE$.

To this aim we now `take the tri-spinor apart' into its 8 + 48 components, 
and write out the correspondence more explicitly for the left-chiral particles
\bea\label{SM1}
U_L^{\bfa\bsa}{} &\equiv&  (u^\bfa, c^\bfa ,t^\bfa)_L \equiv 
\big(\vp^{\bfa \bbE\bV}, \vp^{\bfa \bbZ\bV},\vp^{\bfa\bbD \bV}\big) \nn\\[2mm]
D_L^{\bfa \sa} &\equiv&  (d^\bfa, s^\bfa , b^\bfa)_L  \equiv
\big( \vp^{\bfa\bbZ\bbD}, \vp^{\bfa\bbD\bbE},\vp^{\bfa\bbE\bbZ}\big)\nn\\[2mm]
N_L^{\bsa} &\equiv& (\nu_e, \nu_\mu , \nu_\tau)_L \equiv
\big(\vp^{\bZ\bD\bbV}, \vp^{\bD\bE\bbV}, \vp^{\bE\bZ\bbV}\big) \nn\\[2mm]
E_L^\sa &\equiv& (e^-, \mu^- ,\tau^-)_L \equiv
\big( \vp^{\bbZ\bbD\bbV} , \vp^{\bbD\bbE\bbV} , \vp^{\bbE\bbZ\bbV} \big)
\eea
and for the left-chiral {\em anti}-particles,
\bea\label{SM2}
\bar U_L^{\bbfa\sa}{} &\equiv&  (\bar u^{\bbfa}, \bar c^{\bbfa} , \bar t^{\bbfa})_L \equiv 
\big(\vp^{\bbfa \bE\bbV}, \vp^{\bbfa \bZ\bbV},\vp^{\bbfa\bD \bbV}\big) \nn\\[2mm]
\bar D_L^{\bbfa \bsa} &\equiv&  (\bar d^{\bbfa}, \bar s^{\bbfa} , \bar b^{\bbfa})_L  \equiv
\big( \vp^{\bbfa\bZ\bD}, \vp^{\bbfa\bD\bE},\vp^{\bbfa\bE\bZ}\big)\nn\\[2mm]
\bar N_L^{\sa} &\equiv& (\bar\nu_e, \bar\nu_\mu , \bar\nu_\tau)_L \equiv
\big(\vp^{\bbZ\bbD\bV}, \vp^{\bbD\bbE\bV}, \vp^{\bbE\bbZ\bV}\big) \nn\\[2mm]
\bar E_L^{\bsa} &\equiv& (e^+, \mu^+ ,\tau^+)_L \equiv
\big( \vp^{\bZ\bD\bV} , \vp^{\bD\bE\bV} , \vp^{\bE\bZ\bV} \big)
\eea
We emphasize again that 
the spinors (\ref{SM2}) are not simply complex conjugates of (\ref{SM1}), 
but independent fields (as would be completely obvious if we had written them as 
right-chiral particles). Thus, we now have color indices $\bfa,\bfb, ...\equiv \bE,\bZ,\bD$ and 
family indices $\sa,\sb,...\equiv \bE,\bZ,\bD$ (or $ ([\bbZ\bbD],[\bbD\bbE],[\bbE\bbZ])$), 
thereby disentangling the diagonal SU(3) subgroup into its factors SU(3)$_c$ and SU(3)$_f$.
Accordingly, we stipulate that SU(3)$_c$ acts on indices $\bfa,\bfb,...$ as 
$
D^{\bfa\sa} \ra \ri \la^\bfa{}_\bfb D^{\bfb\bsa}  \,,\,
U^{\bfa\bsa} \ra \ri \la^{\bfa}{}_\bfb U^{\bfb\bsa}
$
where $\la$ is any of the (hermitean) Gell-Mann matrices. Similarly, the SU(3)$_f$ acts on
the indices $\sa,\sb,...$, such that for instance
$
D^{\bfa\sa} \ra \ri \la^\sa{}_\sb D^{\bfa\sb}$ and 
$U^{\bfa\bsa} \ra -\ri (\la^*)^{\bsa}{}_{\bsb} U^{\bfa\bsb}$.
The indices $\bV$ and $\bbV$ are unaffected by these SU(3) actions.

For the electroweak SU(2)$_w$ we treat $(U,D)_L$ and $(N,E)_L$ as doublets, {\it viz.}
{\small
\beq
\begin{pmatrix}  U^{\bfa\bsa} \\ D^{\bfa\sa}  \end{pmatrix}   \ra
\ri \bom\cdot\btau \begin{pmatrix}  U^{\bfa\bsa} \\ D^{\bfa\sa}  \end{pmatrix}  
\;,\quad
\begin{pmatrix}  N^{\bsa} \\ E^{\sa}  \end{pmatrix}   \ra
 \ri\bom\cdot\btau           \begin{pmatrix} N^{\bsa}  \\ E^{\sa} \end{pmatrix}\, , \;                      
\eeq}
with the Pauli isospin matrices $\btau \equiv (\tau^1, \tau^2 ,\tau^3)$
and isospin parameters $\bom$. By contrast, the left-chiral anti-particles 
$(\bar U,\bar D,\bar N,\bar E)$ are treated as SU(2)$_w$ {\em singlets}, and this 
is perfectly consistent because these spinors are independent, as we already said. There is 
no need to discuss U(1)$_Y$ separately, as the associated hypercharges can be 
recovered from the standard formula $Q=T^3 +  Y$. It is an unaccustomed
feature that the electroweak SU(2)$_w$ does {\em not} commute 
with SU(3)$_f$, as the upper and lower components of the electroweak 
doublets are assigned to opposite representations of SU(3)$_f$, such that the
SU(3)$_f$ is realized via the block matrix
\beq
        \begin{pmatrix}  ({\cU}^*)^{\bsa}{}_{\bsb} & 0 \\ 
        0 & {\cU}^\sa{}_\sb \end{pmatrix} \;\; , \quad
       \cU \in {\rm SU(3)}_f
\eeq
in the electroweak isospin space. This prescription is reminiscent of old attempts 
to merge the spin rotation group SU(2) with the flavor symmetry SU(3) into SU(6), but
the groups SU(2)$_w$ and SU(3)$_f$ are expected to be simultaneously realized as
non-commuting symmetries only in the phase with unbroken $\KE$, which does not
admit a local realization within space-time based quantum field theory (this may explain 
why attempts at a group theoretical family unification have not met with success so far).
We recall that the action of $\E$ on the bosonic fields likewise does not
admit a local realization \cite{DHN}. 

While obviously motivated by the known symmetry assignments of quarks and leptons 
the above construction does not tell us how these symmetries act on the gravitinos 
and Goldstinos, and more specifically, whether the SU(3) acting on them should be
identified with SU(3)$_c$ or SU(3)$_f$. It is clear that for consistency this action must be 
the same on (\ref{Gravitino1}) and (\ref{Goldstino}). It  then appears reasonable to 
demand the absence of gauge anomalies for quarks and leptons to persist with 
these fields. The simplest choice ensuring this is to assign (\ref{Gravitino1}) and 
(\ref{Goldstino}) to transform as $\bD\oplus\bbD\oplus\bE\oplus\bbE$ of SU(3)$_c$, 
but as singlets under electroweak SU(2)$_w$ and SU(3)$_f$.  With this choice we 
are led to predict {\em fractionally charged} and {\em strongly interacting} massive 
gravitinos, implying a novel menagerie of exotic Dark Matter candidates, possibly 
even fractionally charged massive bound states of gravitinos and quarks.
Otherwise, the gravitinos would have to be  assigned to chiral representations of 
SU(2)$_w$ (in which case the largest anomaly-free subgroup acting chirally on the 
gravitinos is indeed  SU(2)$\,\times\,$U(1) \cite{Der}) and could acquire 
masses only jointly with electroweak symmetry breaking.

Having realized the action of all SM symmetries on all fermions, we can now
express these actions directly in terms of the original tri-spinor $\chi^{ijk}$, 
that is, in the form
\beq
\chi^{ijk} \ra M^{ijk}{}_{lmn}({\rm G}) \chi^{lmn}    
\eeq
with
\beq
M(\rG)  = \rS^{-1} \circ \rG \circ \rS \, \in \, {\rm SU}(56)
\eeq
where ${\rm G} \,\in\, $ SU(3)$_c \,\times$ SU(2)$_w \,\times$ U(1)$_Y$ or $\in$ SU(3)$_f$.
The matrix $M(\rG)$ is real [that is, $M\in$ SO(56)] for the vectorlike rotations in SU(3)$_c\,$,\, 
SU(3)$_f$ and U(1)$_{em}$, and complex (but still unitary) for chiral electroweak rotations; more 
generally, transformations that do not mix barred and unbarred indices are vectorlike,
whereas transformations that do mix them are chiral. As there is no space here to present 
a complete list of 56-by-56 matrices we have collected explicit formulas in \cite{Link}.

Obviously, $\KE$ is humongously larger than any symmetry so far considered for fermion
unification, and it is therefore all the more remarkable that it admits a realization on precisely
48 spin-$\frac12$ fermions and eight massive gravitinos.  This
enlargement beyond SU(8) hinges crucially 
on the presence of timelike {\em imaginary roots} in $\E$ (which have no analog in finite
dimensional Lie groups), and thus directly on the hyperbolicity of $\E$ \cite{Vigano}. 
However, like for the bosonic sector \cite{DHN}, these considerations are limited, at least 
for the time being, to one given spatial point, and therefore the main challenge remains 
to incorporate the spatial dependence, and to explain how $\KE$ `unfolds' in terms of 
bigger and less unfaithful representations to give rise to emergent space-time and gauge 
symmetries.  A realization of the present scheme would evidently require part of $\KE$ 
to become dynamical ({\em e.g.} with composite vector bosons $W^\pm$ and $Z$), in a partial 
realization of a conjecture already made in \cite{CJ} for the R-symmetry SU(8), but now 
for a  much bigger group, and with a very different mechanism for extracting space-time fields.
Similarly, one may ask 
why the theory should pick a vacuum where $\KE$ is broken to the SM symmetries 
in the way described above; we note, however, that the necessity of cancelling SM gauge
anomalies provides an extremely constraining criterion for vacuum selection and 
symmetry breaking. Finally, there is the question how supersymmetry is ultimately 
disposed of: is it actually there and only broken by some clever mechanism, or is 
it altogether absent even at the most fundamental level? Indeed, there is evidence 
that (conventional) supersymmetry is incompatible with $\KE$ precisely because 
of the presence of imaginary roots \cite{CKN}. This could imply that supersymmetry 
in particle physics may remain a chimera, and even in the context of maximal 
$N\!=\!8$ supergravity not more than a secondary manifestation of a deeper and 
more powerful underlying duality symmetry.\\

While much effort will be required to validate the present proposal (assuming there is any 
truth in it), it offers three main advantages in comparison with other schemes for
fermion unification: (1) it can explain the SM fermion spectrum {\em as is}, without 
$N=1$ superpartners or other exotica, (2) it confirms, and is compatible {\em only} 
with the existence of {\em three generations} of SM fermions,  and (3) it can be directly and 
immediately falsified by the detection of any new fundamental spin-$\frac12$ fermion,  
be it via the discovery of low energy $N\eq 1$ supersymmetry or of new sterile fermions. 
Our proposal underlines the potential importance of the infinite-dimensional exceptional 
algebras $\E$ and $\KE$, and shows how these so far elusive structures might supplant
supersymmetry as a guiding principle for unification, and how they might become 
essential for making contact with SM physics.\\[1mm]
\noindent
 {\bf Acknowledgments:} K.A.~Meissner thanks AEI for hospitality and support. The work of 
 H.~Nicolai has received funding from the European Research Council (ERC) under the 
 European Union's Horizon 2020 research and innovation programme (grant agreement 
 No 740209); he is grateful to G.~Dall'Agata, A.~Kleinschmidt, R.~Koehl and A.~Vigano 
 for discussions.

\end{document}